# Size-Dependent Tensile Behavior of Nanocrystalline HfNbTaTiZr High-Entropy Alloy: Roles of Solid-Solution and Short-Range Order

*Yihan Wu[1,*]   Gaosheng Yan[2,*]   Pengfei Yu[1]   Yaohong Suo[1]*

*Wenshan Yu[2,*]*   Shengping Shen[2]

[1] School of Mechanical Engineering and Automation, Fuzhou University, Fuzhou, PR China

[2] State Key Laboratory for Strength and Vibration of Mechanical Structures, School of Aerospace Engineering, Xi'an Jiaotong University, Xi'an, PR China

[*]**Corresponding authors:**

wuyihan@fzu.edu.cn (Y Wu), ygs@xjtu.edu.cn (G Yan), wenshan@mail.xjtu.edu.cn (W Yu)

**Abstract:** This study investigates the size-dependent mechanical behavior of the HfNbTaTiZr refractory high-entropy alloy (RHEA) under uniaxial tension, with a focus on the effects of random solid-solution (RSS) and chemical short-range order (CSRO). A machine learning framework is developed to accelerate the parameterization of interatomic force fields (FFs), enabling molecular dynamics simulations of three nanocrystalline models: (i) a meta-atom (MA) model representing the RHEA as a hypothetical single-element system with averaged properties, (ii) a quinary RSS model with randomly distributed constituent atoms, and (iii) a Monte Carlo (MC) model with internal CSRO. The results reveal that RSS enhances strength, while CSRO reduces flow stress level but improves strain hardening and failure resistance. A transition from Hall–Petch (HP) strengthening to inverse Hall–Petch (IHP) softening is observed, with CSRO suppressing this transition. The underlying plastic mechanisms (i.e., dislocation slip, deformation twinning, phase transformation and grain boundary movements) are analyzed from both nanostructural and energetic perspectives. Theoretical models are established to describe the size-dependent yield strength and predict the critical grain size. Additionally, the contributions of different plastic mechanisms to the overall stress response are separately quantified. These findings provide new insights into the design and performance optimization of RHEAs through nanostructural engineering.

1. **Introduction**

Recently, the refractory high-entropy alloys (RHEAs) have been drawing extensive attentions from both scientific and industrious fields [1, 2]. These alloys typically contain five or more metallic elements mainly from Groups VI ~ VI of the periodic table which are combined in a near-equimolar way, leading to a body-centered-cubic (BCC) solid-solution structure. This novel design concept significantly expands the compositional space of alloy materials, endowing RHEAs four core effects [3-7]: high entropy effect, distorted lattice, sluggish diffusion and cocktail effect. As a result, the RHEAs usually exhibit exceptional properties better than their constituents. Notably, the HfNbTaTiZr quinary alloy distinguishes among these RHEAs owing to its high strength-to-weight ratio and room temperature ductility [8, 9]. It is therefore regarded as a promising candidate to meet the constantly growing performance requirements in various engineering applications including aerospace engineering [6], nuclear powerplant [10] and micro-electromechanical systems [11].

Despite the tempting prospects, research on the HfNbTaTiZr RHEA stays in a nascent stage. Currently, most studies focus on improving fabrication techniques or optimizing chemical compositions to enhance the synergy between strength and ductility [1, 6, 12]. Modulating the performance of HfNbTaTiZr through nanostuctural design has been rarely studied. For conventional metals and alloys, grain size engineering has been widely utilized to improve performance. Nanocrystalline metals exhibit a 'smaller-is-stronger' behavior with decreasing grain size (i.e., the Hall–Petch (HP) strengthening), which is commonly ascribed to dislocation pile-ups at grain boundaries (GBs) in fine-grained materials [13, 14]. When grain size decreases below a critical value, the HP strengthening breaks down and a further grain refinement causes a reduced strength. This inverse HP (IHP) softening results from GB-mediated processes (e.g., migration or sliding) [15, 16]. While this size-dependent approach has successfully guided the design of high-performance conventional alloys, its extended application to RHEAs is inhibited by the complex plastic deformation behaviors. Studies have revealed that multiple plastic mechanisms (e.g., dislocation slip, deformation twinning, phase transformation, etc.) can be concurrently activated in RHEAs, depending on loading conditions and environmental factors [8, 9, 17, 18]. This complexity makes it challenging to experimentally track nanostructural evolution during plastic flow, limiting a comprehensive understanding of size-dependent mechanical properties. Numerical simulations thus offer a viable alternative to this issue.

Some physically based numerical methods, such as crystal plasticity finite element method (CPFEM) and discrete dislocation dynamics (DDD), have been developed to investigate the micro- and nanoscale plasticity of RHEAs. For instance, CPFEM has correctly predicted the strain hardening of HEAs by incorporating contributions from dislocations, GBs[15, 19], twin boundaries (TBs)[20] and second phases[21] into the constitutive laws; results are consistent with experiment data. Unfortunately, CPFEM, as a continuum method, is limited in capturing the random solid-solution (RSS) and chemical short-range order (CSRO) effects induced by complex elemental interactions in RHEAs. These atomic-scale heterogeneities are beyond the capability of constitutive equations relying on homogenized representative material parameters. On the other hand, the DDD method can be utilized to investigate the influence of atomic-scale heterogeneity on the mechanical response[22, 23]. However, this method mainly considers dislocation slip and is thus insufficient in describing the complex plastic mechanisms observed in RHEAs.

Luckily, molecular dynamics (MD) simulations offer a powerful tool to comprehensively study nanoscale plasticity and the underlying nanostructure-property relationships in RHEAs [11, 24]. Nonetheless, their application to HfNbTaTiZr remains limited owing to the lack of reliable interatomic force fields (FFs). Because of the mathematically complex formulation [10, 25, 26] and large number of parameters, developing an FF that accurately reproduces various material properties is computationally demanding, particularly for multicomponent systems like RHEAs. To date, most MD studies using existing FFs focus on the CoNiFe-based HEAs with face-centered-cubic (FCC) structures. While these studies partly explain the size-dependent mechanical properties of FCC HEAs through competitions among different plastic mechanisms [27-30], a key limitation persists: there is still no effective approach to separate and individually analyze the RSS and CSRO effects. The multi-element atomic configurations used in simulations always concurrently contain these two effects. Although some studies have compared the mechanical properties of HEAs with pure metals [31], the RSS effect is still much unclarified due to the disparate properties of HEAs and their constituents. Recently, Wang *et. al.*[32] proposed the meta-atom (MA) method which assumes that mechanical properties of an alloy are primarily governed by a finite set of material constants. Thus, the complex alloy system can be approximated by a hypothetical pure metal that exhibits 'average' properties of the alloy [18, 32, 33]. This approach suggests a novel means to effectively decouple RSS and CSRO effects, by comparing the mechanical responses of the MA surrogate system with the full-element system.

Based on current research status, some critical questions arise regarding the size-dependent nanoscale mechanical properties and behaviors of HfNbTaTiZr RHEA:

1. Is there a feasible way to accelerate the development of FFs for complex alloy systems? How to establish an applicable FF for the HfNbTaTiZr RHEA that accurately predicts its mechanical properties and nanoscale deformation behaviors?

2. How do the mechanical properties and plastic mechanisms of HfNbTaTiZr vary with grain size? How to separately analyze the effects of RSS and CSRO on such a size-dependence?

3. How to theoretically describe the size-dependent mechanical properties and predict the optimal grain size? How to quantify contributions from different plastic mechanisms?

In this study, we attempt to address these questions by first developing a machine learning (ML) accelerated workflow for FF parameterization. Using this workflow, we then construct alloy FF and MA FF to study the RSS and CSRO effects on the size-dependent mechanical properties and nanoscale plastic mechanisms. Finally, based on simulation data, theoretical models are proposed to describe the size-dependent yield strength; the contributions of different deformation mechanisms to the stress-strain response are also quantified. Our findings provide guidance for the design and performance optimization of HfNbTaTiZr RHEA.

**2. Computational details**

## 2.1 Parameterization of interatomic force field (FF)

The reliability of molecular dynamics (MD) simulations depends on the accuracy of the adopted force field (FF). The FF directly defines atomic interactions based on spatial arrangements and determines the evolution trajectory of the particle system [10, 25]. Despite its importance, FF training remains challenging owing to the complex mathematical form and numerous internal parameters. These parameters must be carefully determined to minimize deviations between FF predictions and *ab initio* reference data, typically via certain optimization algorithms [26]. Each optimization iteration requires a series of independent MD runs to update current FF predictions for key materials properties (e.g., lattice parameters, cohesive energies, elastic constants, etc.), which is expensive both in time and computational resources.

To enhance the efficiency of FF parameterization, we introduce a machine learning (ML) strategy. As illustrated in Fig. 1a, a surrogate model is trained to directly generate material properties based on given FF parameters, thus bypassing the time-consuming MD calculations. Here, the Zhou's form of embedded atom method (EAM) is selected [10, 34]. It contains 14 independent parameters for each metal constituent, which serve as input variables for the ML model. A back-propagation neural network (BPNN) is constructed to learn the relationship between FF parameters and material properties (detailed in Table 1). To ensure consistency in output units (i.e., eV), the BPNN predicts the energies of deformed atomic configurations and derives lattice and elastic constants (in Å and GPa) through energy-volume relationships [35]. The BPNN, with a batch size of 1,024, contains two hidden layers and 50 nodes in each layer (Fig. 1b). The exponential linear unit (ELU) activation function and the Adam algorithm are employed. In order to prevent overfitting, the initial learning rate of $10^{-3}$ is gradually reduced via cosine annealing; an early-stopping criterion terminates training when the mean absolute error (MAE) ceases to decrease for 300 epochs. The training dataset, comprising of 18,592 data points, is obtained from our previous works of FF development [10, 36]. The dataset is split into 90% for training and 10% for testing. The final performance of the BPNN is shown in Fig. 1c.

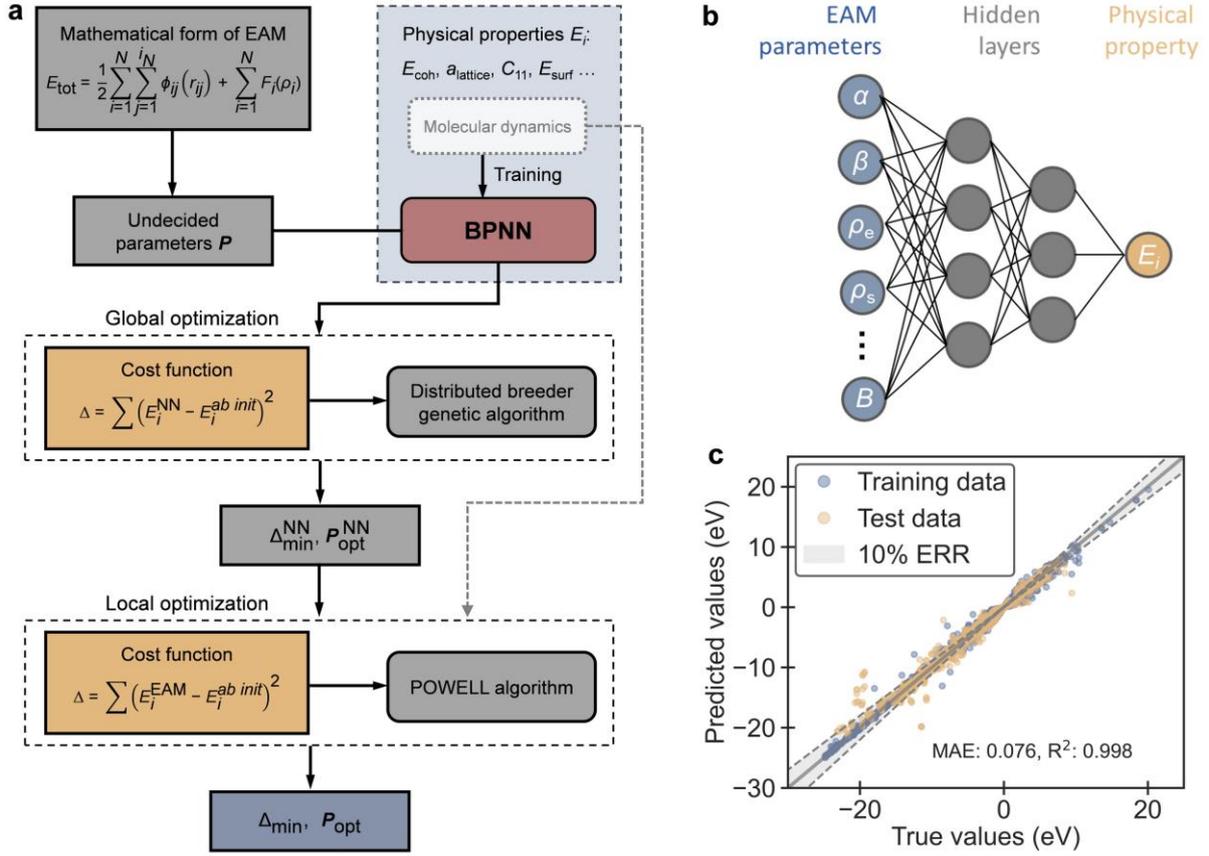

**Fig. 1** (a) A flowchart showing the parametrization process of the Hf/Nb/Ta/Ti/Zr EAM FF. (b) The schematic configuration of the BPNN that predicts material properties based on FF parameters. (c) Accuracy of the BPNN surrogate model after training.

**Table 1** The physical properties as the target variables predicted by the BPNN. The mechanical constants and lattice parameter are derived from energies of deformed configurations via energy-volume relationships.

| Properties | Description |
|---|---|
| $E_{coh}$ | Cohesive energies of BCC, FCC and HCP crystals at equilibrium |
| $a$ | Lattice parameters of BCC, FCC and HCP crystals |
| $C_{ij}$ | Elastic constants of the ground-state crystalline structure |
| $B$ | Bulk modulus for BCC, FCC and HCP structures |
| $E_v^f$ | Vacancy formation energy of the ground-state crystalline structure |
| $E_{surf}$ | Surface energy of the ground-state crystalline structure |
| $E_{GSFE}$ | Generalized stacking fault energy |

Using the trained BPNN, the EAM FF parameterization process is significantly accelerated. As outlined in Fig. 1a, the surrogate model can instantly return material properties for a given parameter set $P$. The cost function $\Delta$, defined as the sum of squared deviations between BPNN predictions and *ab initio* data, is then minimized using a distributed breeder genetic algorithm (DBGA) [37]. This global optimizer is reported to outperform conventional genetic algorithms [10, 25]. After DBGA convergence, the resulting parameter set $\boldsymbol{P}_{\text{opt}}^{\text{NN}}$ corresponds to the global minimum on the parameter hypersurface reconstructed by BPNN. Subsequently, an additional local optimization using the POWELL algorithm [38] is performed, during which MD simulations (instead of BPNN outputs) are used to evaluate Δ. This step corrects any inaccuracies introduced by the surrogate model. Due to the high accuracy of the BPNN (Fig. 1c), this local optimization can converge within a few iterations and therefore impart little influence on the overall time efficiency.

Following this workflow, we can easily develop FFs for the five pure metals Hf, Nb, Ta, Ti and Zr. The alloy interactions between metallic constituents are modeled using the widely accepted empirical mixing law [39]

$$\phi^{ab} = \frac{1}{2}\left(\frac{f^b(r)}{f^a(r)}\phi^{aa}(r) + \frac{f^a(r)}{f^b(r)}\phi^{bb}(r)\right),\quad (1)$$

where $\phi$ denotes pair interactions; $a$ and $b$ represent different chemical species. This mixed FF is hereafter referred to as the 'alloy FF'. Moreover, inspired by the meta-atom (MA) theory [18, 32], we also develop a special FF for the HfNbTaTiZr RHEA, representing it as a hypothetical element with averaged properties of the alloy (namely, the 'MA FF'). Using the workflow in Fig. 1a, both the alloy and MA FFs can be developed within a few hours, offering substantial speedup compared to traditional FF training methods

For details on the mathematical formulation of EAM, final FF parameters for the RHEA, and the cross-validation procedure of FFs, readers can refer to section S1 of the supplementary material.

## 2.2 Simulation methodology

To study the grain size effect on the mechanical properties and plastic behaviors of HfNbTaTiZr RHEA, BCC polycrystalline configurations are generated through Voronoi polyhedral tessellation using ATOMSK software [40] (Fig. 2a). The configurations, whose sizes are listed in Table 2, contain ~698,000 to ~6,854,200 atoms. Periodic boundary conditions are imposed in all directions. Varying grain sizes $D$ from 4.0 nm to 25.0 nm are investigated. For configurations with fewer than 20 grains (i.e., $D > 10.0$ nm), three independent samples are generated to account for statistical variations in grain orientations.

To isolate the effects of random solid-solution (RSS) and chemical short-range order (CSRO), different chemical environments are modeled in the configurations. The first set of configurations consists of a single hypothetical chemical species, with properties (e.g., cohesive energy, lattice constant, elastic constants and defect energetics) identical to averaged values of the HfNbTaTiZr RHEA. Hereafter, these models are referred to as meta-atom (MA) models [18, 32]. The second type includes all five elements (Hf, Nb, Ta, Ti and Zr) randomly distributed in the nanocrystals. These models (hereafter denoted as RSS models) allow us to investigate the RSS effect. The third type (i.e., MC models) is created by applying Monte Carlo (MC) optimization to the RSS models to introduce local elemental ordering. Fig. 2b displays the Warren-Cowley CSRO parameters [10, 25, 41] in the MC models for various elemental pairs. Negative values indicate clustering tendencies among Hf-Zr-Ti and Nb-Ta. Higher peaks in the pair distribution functions (Fig. 2c) also suggest clustering of these element pairs. Such a phenomenon is consistent with previous experimental observations [42, 43] and validate the developed FF. The MC models allow us to assess the impact of CSRO on RHEA mechanical properties.

All models (MA, RSS and MC) undergo energy minimization via conjugate gradient method, followed by 200 ps equilibration at room temperature in the NPT ensemble to release internal stress. Thereafter, uniaxial tension is applied along the $z$-direction at a constant strain rate of $10^8$/s. This strain rate is commonly adopted in MD simulations because previous studies have reported that strain rates below $5\times10^9$/s yield reliable results consistent with experimental observations [11, 44, 45]. During loading, boundaries in the $x$ and $y$ directions are set free to contract, corresponding to plane-stress conditions.

All simulations use a timestep of 1.0 fs and maintain temperature at 300 K via the Nosé-Hoover thermostat [46]. The LAMMPS software [47] is used for simulations, and results are visualized using OVITO [48]. The atomic structures are analyzed using the Crystal Analysis Tool (CAT) package [49].

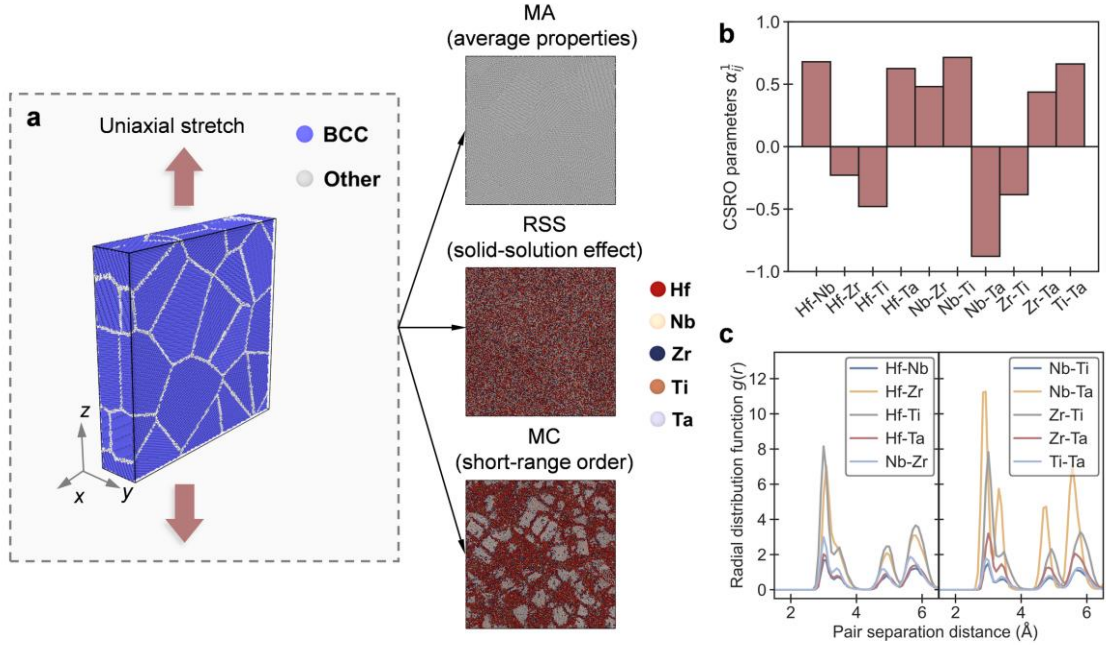

**Fig. 2** (a) Schematic of the nanocrystalline MA, RSS and MC models under uniaxial stretch. (b) The CSRO parameter of the MC models. (c) The pair distribution functions of different element pairs in the MC models.

**Table 2** Polycrystalline RHEA configurations constructed in this study. Considering the polyhedral shaped grains, the average grain size is approximated by $D = (l_x \times l_y \times l_z / N)^{1/3}$, where $l_x$, $l_y$, $l_z$ are model dimensions and $N$ is the number of grains.

| Dimensions (nm³) | Number of grains | Average grain size (nm) | Number of atoms | Number of configurations |
|---|---|---|---|---|
| 40×8×40 | 200 | 4.00 | ~698,000 | 1 |
| 40×8×40 | 59 | 6.01 | ~698,000 | 1 |
| 40×8×40 | 25 | 8.00 | ~698,000 | 1 |
| 40×10×40 | 16 | 10.00 | ~872,600 | 3 |
| 40×12×40 | 11 | 12.04 | ~1,047,200 | 3 |
| 40×14×40 | 8 | 14.09 | ~1,221,700 | 3 |
| 40×16×40 | 6 | 16.22 | ~1,396,300 | 3 |
| 60×20×60 | 9 | 20.00 | ~3,927,000 | 3 |
| 75×25×75 | 9 | 25.00 | ~6,854,200 | 3 |

## 3. Results

## 3.1 Materials properties of the HfNbTaTiZr alloy

Fig. 3 displays some important material properties of the HfNbTaTiZr RHEA. It is apparent that both the alloy FF and the MA FF accurately reproduce the cohesive energy, lattice parameter and elastic constants in well consistence with reference values calculated via *ab initio* density functional theory (DFT). Fig. 3 also demonstrates that the properties of RHEA can be described using those of its constituent metals through the rule of mixture (ROM), expressed as [9]

$$Q_{RHEA} = \sum_i c_i Q_i, \qquad (2)$$

where $Q$ represents a specific physical property, $c$ is the chemical concentration and $i$=1~5 denotes the five constituent metals. Besides, Fig. 4 presents the cohesive energies of BCC, FCC and HCP crystals calculated using DFT, ROM, alloy FF and MA FF, respectively. Clearly, the developed FFs correctly predict the energy differences among these phases, which basically comply with ROM. The small energy differences between phases imply that phase transformation is probably an important plastic deformation mechanism of the RHEA. Together, Figs. 3 and 4 confirm the accuracy and applicability of the developed FFs. More detailed validation procedures are provided in section S1.2 of the supplementary material.

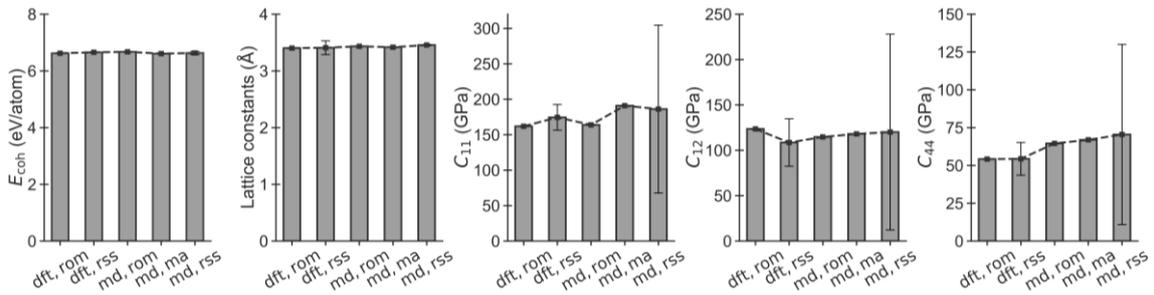

**Fig. 3** Important properties (cohesive energy $E_{coh}$, lattice parameter $a$ and elastic constants $C_{11}$~$C_{44}$) of the RHEA calculated using DFT and MD methods. Properties with the 'rss' subscript are calculated using the equiatomic RSS model; those with the 'ma' subscript are results based on the MA model; those with the 'rom' subscript are calculated through the rule of mixture (i.e., average properties of constituent metals).

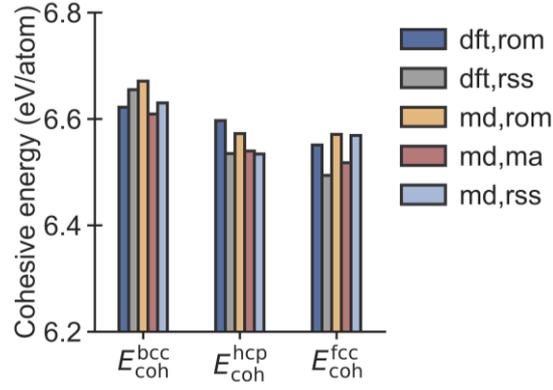

**Fig. 4** The energy differences between BCC, FCC and HCP crystalline phases of the HfNbTaTiZr RHEA.

Moreover, it is usually accepted in theoretical modelling that mechanical behaviors of alloys are dictated by only a few key material properties [18, 32, 33]. Alloys with identical material constants are expected to exhibit similar mechanical responses. According to this assumption, the similar energetic, structural and mechanical constants of the MA and RSS models (Figs. 3 and 4) suggest that alloying effects on the deformation behavior of HfNbTaTiZr are limited, and its mechanical properties should equal the average of its constituents. However, this obviously contradicts recent experiments revealing 'cocktail' effects in RHEAs [4-7]. The following sections aim to explain this discrepancy and further clarify the alloying effects (including RSS and CSRO effects) on the mechanical properties and plastic mechanisms of the RHEA.

**3.2 Grain size effect on mechanical properties**

This section discusses the grain size dependence of the modulus $E$, yield strength $\sigma_Y$, ultimate strength $\sigma_U$ and average flow stress $\sigma_F$ of the HfNbTaTiZr RHEA. Definitions of these mechanical properties are given in Table 3. Fig. 5 displays the variation of these properties for MA, RSS and MC nanocrystals as a function of the inverse square root of grain size $D^{-1/2}$ (stress-strain curves for varying grain sizes are shown in section S2 of the supplementary material). Apparently, the RSS RHEAs (Figs. 5a~5d) exhibit higher modulus, strengths and flow stress compared with the MA RHEAs. Such a phenomenon indicates that the RSS effect significantly improves overall mechanical properties. On the other hand, Figs. 5a~5d also suggest that the presence of CSRO is detrimental to the mechanical properties of the quinary RHEA, as evidenced by the noticeable reduction in these values. Through simulations using nanocrystalline RHEAs with different internal chemical environments, the effects of RSS and CSRO can be separated, revealing opposite influences on mechanical properties. The mechanism underlying these opposite effects are explored in section 3.3, through analyses on nanostructural evolution.

Table 3 Definition of mechanical properties

| Mechanical property | Definition |
|---|---|
| Modulus $E$ (GPa) | Slop of the stress-strain curve up to 2.5% strain |
| Yield strength $\sigma_Y$ (GPa) | 0.2% offset yield stress |
| Ultimate strength $\sigma_U$ (GPa) | Maximum of the stress-strain curve |
| Flow stress $\sigma_F$ (GPa) | Average stress during the steady-state flow stage (i.e., 10% ~ 20% strain) |

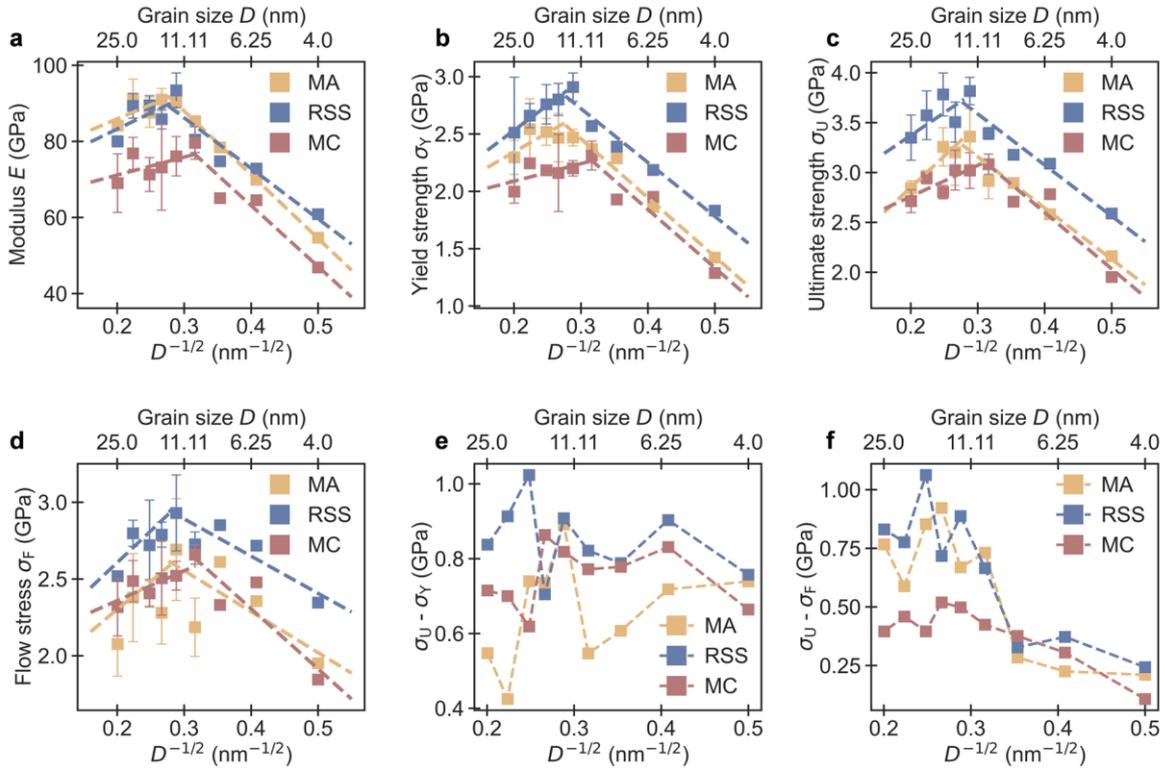

**Fig. 5** (a)~(d) The modulus $E$, yield strength $\sigma_Y$, ultimate strength $\sigma_U$ and average flow stress $\sigma_F$ of MA, RSS and MC RHEA nanocrystals vs. grain size. (e)~(f) Values of $\sigma_U - \sigma_Y$ and $\sigma_U - \sigma_F$ as functions of grain size $D$.

Figs. 5a~5d also reveal that as grain size decreases, the mechanical properties initially increase, corresponding to a HP strengthening regime. This variation trend then reverses after $D$ drops below a certain value $D_c$ around 10.0 nm, consistent with previous studies of other HEAs [11, 28, 50]. This reversed trend marks the onset of the IHP softening stage. Both the HP and IHP

regimes can be depicted by the following expression

$$X = X_0 + kD^{-1/2} \tag{3}$$

where $X$ represents a specific property ($E$, $\sigma_Y$, $\sigma_U$ or $\sigma_F$), $X_0$ is the property of a single crystal (i.e., $D \rightarrow \infty$), and $k$ is a material-dependent constant. The fitted values of $k$ and $X_0$ as well as the critical grain size $D_c$ are listed in Table 4. Interestingly, the MA and RSS models show similar $D_c$ (~13 nm), whereas the MC models have a smaller $D_c$ (~10 nm). This discrepancy in $D_c$ suggests that internal CSRO suppresses the transition of tensile response from HP strengthening to IHP softening, while the RSS effect has minimal influence on such a transition. In addition, similar $k$ values for MA, RSS and MC models during the IHP stage (Fig. 5 and Table 4) suggest that neither RSS nor CSRO significantly affects the softening rate as $D$ decreases. In contrast, the MC models exhibit notably lower $k$ values in the HP stage compared to MA and RSS models, indicating that CSRO decelerates the strengthening effect with decreasing $D$ in this stage.

**Table 4** Fitting parameters in Eq. 3 for MA, RSS and MC nanocrystalline RHEAs

| Properties | Stage | MA | | | RSS | | | MC | | |
|---|---|---|---|---|---|---|---|---|---|---|
| | | $k$ (GPa·nm$^{1/2}$) | $X_0$ (GPa) | $D_c$ (nm) | $k$ (GPa·nm$^{1/2}$) | $X_0$ (GPa) | $D_c$ (nm) | $k$ (GPa·nm$^{1/2}$) | $X_0$ (GPa) | $D_c$ (nm) |
| $E$ (GPa) | HP | 82.84 | 68.57 | 12.92 | 84.83 | 66.26 | 13.26 | 47.30 | 61.66 | 10.02 |
| | IHP | -167.41 | 138.20 | | -132.80 | 126.02 | | -160.31 | 127.26 | |
| $\sigma_Y$ (GPa) | HP | 3.41 | 1.66 | 13.34 | 4.38 | 1.66 | 13.45 | 1.53 | 1.78 | 9.98 |
| | IHP | -5.14 | 4.00 | | -4.70 | 4.13 | | -5.09 | 3.88 | |
| $\sigma_U$ (GPa) | HP | 6.38 | 1.57 | 13.49 | 4.55 | 2.46 | 13.20 | 3.12 | 2.14 | 10.35 |
| | IHP | -5.14 | 4.70 | | -5.12 | 5.12 | | -5.65 | 4.86 | |
| $\sigma_F$ (GPa) | HP | 3.71 | 1.56 | 12.59 | 4.12 | 1.78 | 12.71 | 1.94 | 1.97 | 9.48 |
| | IHP | -2.67 | 3.36 | | -2.42 | 3.62 | | -3.92 | 3.88 | |

To further assess the mechanical properties, we calculate the values of $\sigma_U - \sigma_Y$ and $\sigma_U - \sigma_F$. The former represents strain hardening ability after yielding, while the latter relates to failure resistance. Fig. 5e manifests that both RSS and CSRO effects enhance strain hardening, as evidenced by increased values of $\sigma_U - \sigma_Y$ in the RSS and MC models. Meanwhile, MA nanocrystals with grain sizes near $D_c$ also exhibit improved hardening ability. On the other hand, all models show low $\sigma_U - \sigma_F$ values during the IHP stage, denoting strong failure resistance by maintaining stress levels after peak load. However, this failure resistance rapidly diminishes in the MA and RSS models once $D > D_c$ (the HP stage), whereas the MC models maintain failure resistance across a broader grain size range. These findings imply that despite relatively reduced modulus, strengths and flow stress, internal CSRO can enhance both strain hardening and failure resistance.

### 3.3 Nanoscale plastic deformation behavior

In this section, the nanoscale plastic deformation behaviors of MA, RSS and MC nanocrystalline RHEAs are analyzed by inspecting nanostructural evolution during uniaxial stretch. The nanocrystals with large ($D = 20$ nm) and small ($D = 4$ nm) grain sizes are selected to elucidate the deformation mechanisms in the HP and IHP regimes, respectively. Fig. 6a presents the stress-strain response of an MA nanocrystal with $D = 20$ nm, accompanied by snapshots of atomic structures at various strain levels (Figs. 6b~6e). At the elastic stage, uniform elongation occurs and a high density of dislocations is observed at grain boundaries (GBs). Such an entangled dislocation network stores strain energy and induces stress concentration at GBs, as reported by a previous MD study [11]. At tensile strain $\varepsilon=3.6\%$, these dislocations emit from GBs into the grain interiors (Figs. 6b1 and 6b2), activating yielding of the nanocrystal. The emitted dislocations are mainly 1/2<111> types with both edge and screw characters (Fig. S5 of the supplementary material). This phenomenon is consistent with previous experiments showing increased edge dislocation activity in BCC RHEAs relative to conventional BCC metals [51, 52]. During subsequent plastic flow, the emitted dislocations either interact within the grains, or glide until annihilation in GBs at the opposite side of the grain. The former leads to formation of sessile dislocations that help to maintain the load capacity of the grains [53], while the latter causes strain energy release and stress relaxation. Besides, emission of dislocations also induces mass transport that carries away atoms at GBs, leading to coalescence of adjacent GBs as indicated by Fig. 6e3. Such a phenomenon is also common during loading activated growth of nanovoids in metals [53-55]. In addition, we observe that 1/2<111> dislocations can dissociate into three 1/6<111> partials on successive atomic planes, which results in a 3-layer growth behavior of deformation twins (Figs. 6c1 and 6c3), in agreement with prior reports for BCC metals [56, 57].

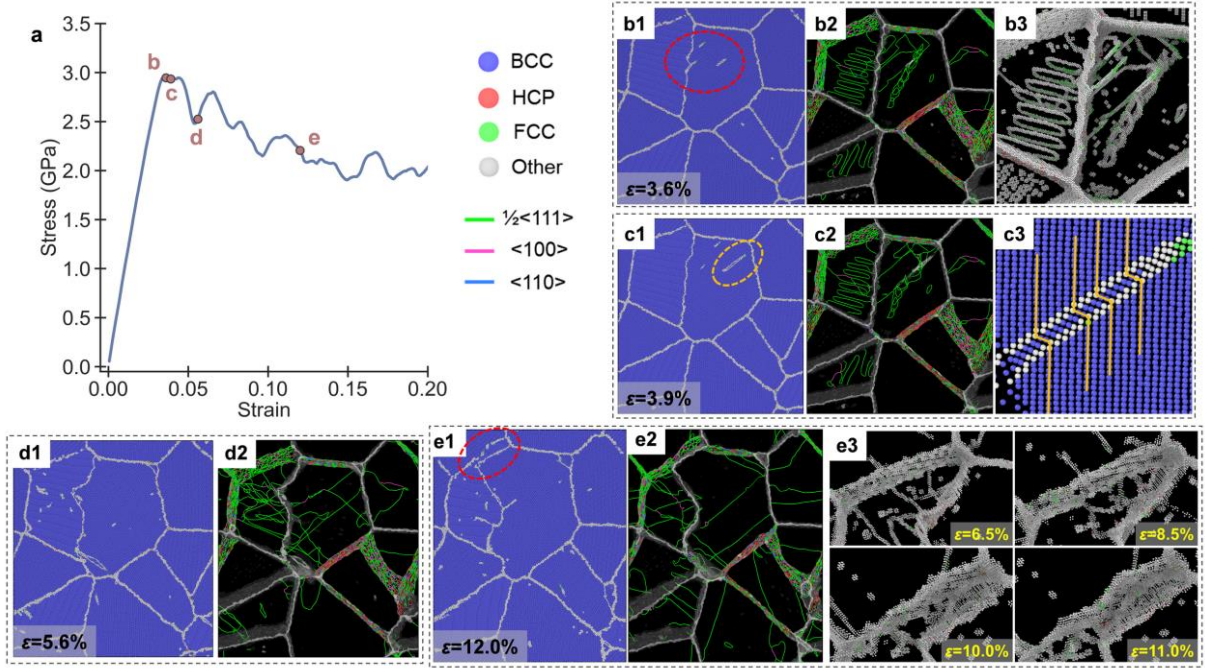

**Fig. 6** (a) The stress-strain curve of an MA nanocrystal with $D$ = 20 nm. (b1)~(e1) Atomic structures at different strains marked in (a). Atoms are colored according to their structural type. Red and yellow dashed circles mark intragrain dislocations and deformation twins, respectively. (b2)~(e2) The 3D-distribution of GBs, TBs and dislocations. (b3)~(e3) Enlarged images of circularly marked regions in (b1)~(e1).

To examine the RSS effect on the plastic deformation behavior, Fig. 7 displays the tensile response and structural evolution of an RSS nanocrystal with $D$ = 20 nm. A higher stress level (Fig. 7a) relative to MA counterpart (Fig. 6a) indicates RSS enhanced mechanical properties, which is attributed to alterations in nanoscale deformation behavior. In addition to dislocation emission from GBs, the initial plasticity in the RSS nanocrystal is also marked by a BCC-to-FCC phase transformation (Fig. 7b3). Notably, this transformation is partly reversible as indicated in Figs. 7b3~7d3; a recovery to the BCC phase occurs once alternative plastic mechanisms (e.g., dislocation slip and twinning) release the internal strain energy. Similar transient BCC-to-FCC transformation is also reported for other polycrystalline RHEAs [4, 10]. Moreover, a remarkable increase in the density of dislocations and twin boundaries (TBs) is observed during plastic deformation, which are the primary factors to the improved mechanical property. The multicomponent nature of RSS nanocrystals creates a complex local activation energy landscape [33, 58, 59], facilitating dislocation activities such as multiplication, pinning and reaction (Figs.

7b2~7e2). As a result, strain hardening is enhanced so that strengths and flow stress of the RSS nanocrystal are elevated. On the other hand, the higher dislocation density also raises the frequency of dislocation dissociation into twin partials, promoting deformation twinning and the formation of nanotwin lamellae (Figs. 7c3~7e3) as a widely accepted strengthening mechanism [11, 60].

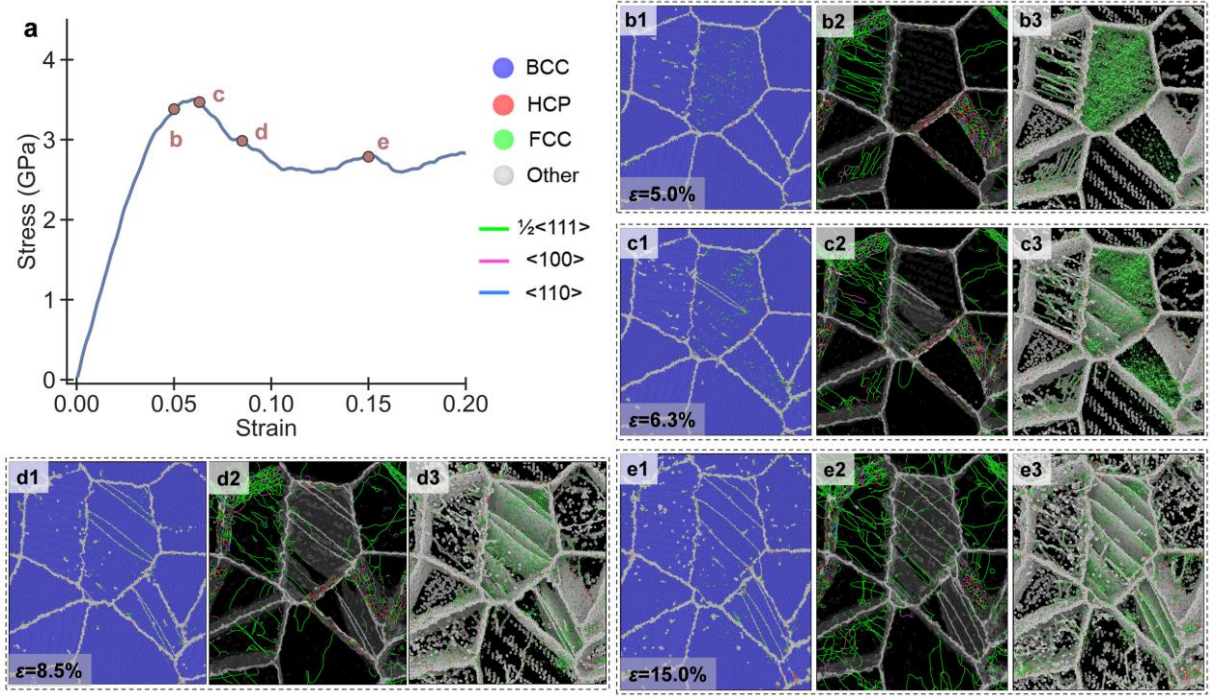

**Fig. 7** (a) The stress-strain curve of an RSS RHEA nanocrystal with $D = 20$ nm. (b1)~(e1) Atomic structures at different strains marked in (a). (b2)~(e2) The 3D-distribution of GBs, TBs and dislocations. (b3)~(e3) Deformed structures after removal of all BCC atoms.

For the MC nanocrystal with $D = 20$ nm, the internal CSRO greatly reduces the stress level, as presented in Fig. 8a1. The decomposition of an ideal solid-solution phase into Hf-Zr-Ti-rich GB regions and Nb-Ta-rich grain interiors is evident (Figs. 8a2~8a4). As a result, Hf, Zr and Ti atoms at GBs in the undeformed MC nanocrystal undergo spontaneous martensitic transformation to their ground-state HCP structure (Figs. 8a3 and 8a4). During the subsequent tensile deformation, the GB martensite acts as nucleation sites for continuous phase transformation as revealed by Figs. 8b3~8e3. As strain increases, BCC grains are progressively consumed by close-packed FCC and HCP phases. Previous studies [60, 61] have reported that transformation induced plasticity (TRIP) often enhances ductility but lowers strength compared to twinning induced plasticity (TWIP), consistent with Figs. 5 and 8. Therefore, CSRO facilitates elemental clustering, leading to TRIP-dominated plastic mechanism and reduced stress levels during tensile loading.

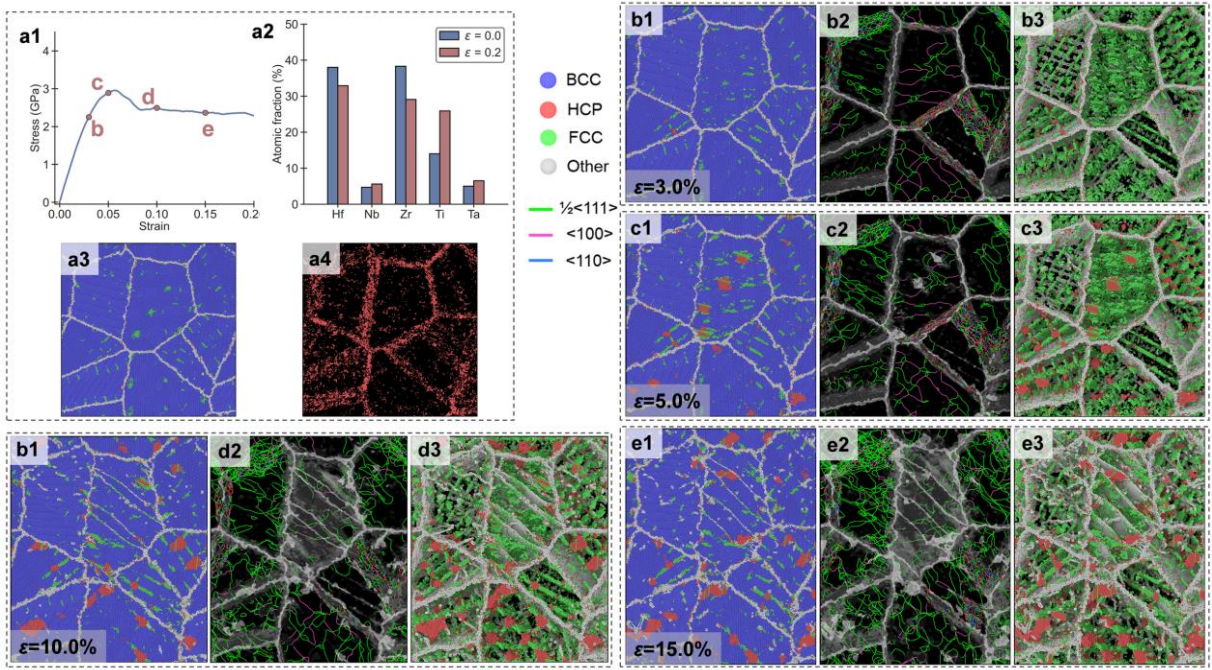

**Fig. 8** (a1) The stress-strain curve of an MC RHEA nanocrystal with $D$ = 20 nm; (a2) the composition of the HCP phase at strains 0.0 and 0.2; (a3) the undeformed nanocrystal and (a4) the associated distribution of HCP atoms. (b1)~(e1) Atomic structures at different strains marked in (a1). (b2)~(e2) The 3D-distribution of GBs, TBs and dislocations. (b3)~(e3) Deformed structures after removing all BCC atoms.

When grain size drops below the critical value $D_c$, significant changes in plastic behavior occur, which causes the IHP softening. Fig. 9 presents the stress-strain response and structural evolution of an MA nanocrystal with $D$ = 4 nm. The density of GB dislocations ($\rho_{GB}$= 0.068 nm$^{-2}$ in undeformed nanocrystal) is notably lower than in the nanocrystal with $D$ = 20 nm (Fig. 6, $\rho_{GB}$= 0.131 nm$^{-2}$ ). The shortened dislocation lines cannot entangle and form a network to stabilize GBs. Consequently, the plastic flow is dominated by GB motion, namely, migration of GBs. For example, as tensile strain increases from 1.0% to 6.0%, active GB migration (Figs. 9b3 and 9c3) leads to grain growth (grains 1 and 4), grain disappearance (grains 2, 3, 5 and 7), and the emergence of new grains (grains 8 and 9). This GB motion induces a noticeable drop of stress level, which explains the IHP softening. Besides, such GB-dominated plasticity enhances failure resistance, evidenced by the suppressed stress relaxation after the ultimate strength is reached (Fig. 5f).

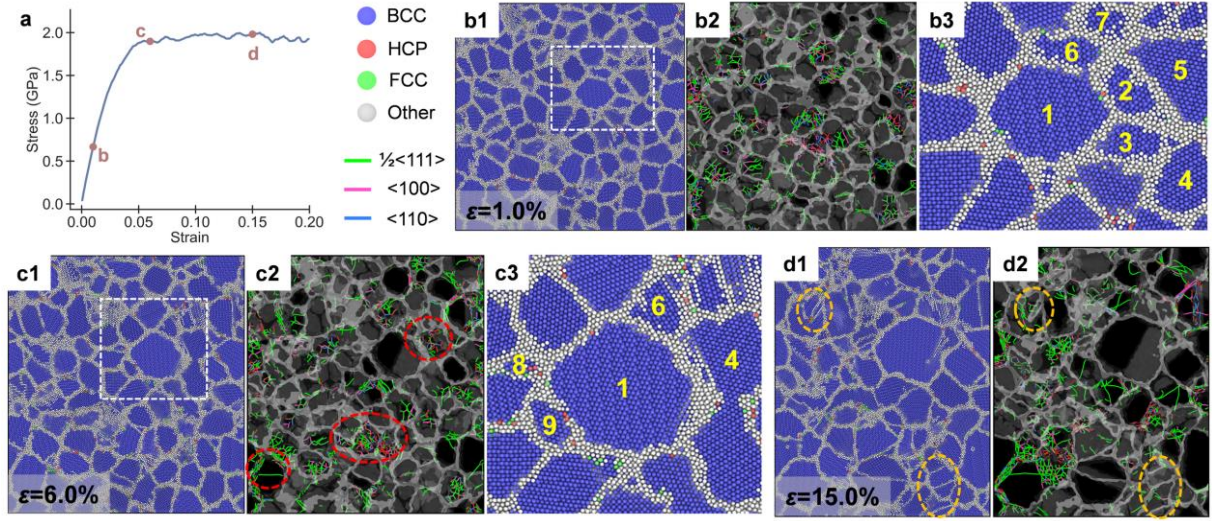

**Fig. 9** (a) The stress-strain curve of an MA RHEA nanocrystal with $D = 4$ nm. (b1)~(d1) Atomic structures at different strains marked in (a). (b2)~(d2) The 3D-distribution of GBs, TBs and dislocations lines. (b3) and (c3) are enlarged images of rectangularly marked regions in (b1) and (c1). Red and yellow dashed circles mark intragrain dislocations and deformation twins, respectively.

For the RSS nanocrystal with $D = 4$ nm (Fig. 10), the plastic behavior differs from the HP regime (Fig. 7). Specifically, dislocation activities are suppressed as evidenced by reduced dislocation density (Figs. 10b2~10d2). Likewise, the phase transformation is considerably suppressed. Notably, deformation twinning still occurs as manifested in Fig. 10d, contributing to the highest stress level in comparison to the MA and MC models at the same grain size. However, the plastic flow is governed by GB sliding and amorphization (Fig. 10b1~10d1), rather than GB migration as in the MA nanocrystal (Fig. 9). Previous studies have demonstrated that multiple alloy elements induce strongly distorted structural units of GBs [62, 63], promoting a transition from GB migration to sliding and amorphization. A similar plastic behavior dominated by GB sliding and amorphization is also observed in the MC nanocrystal with $D = 4$ nm (Fig. 11). No obvious grain growth occurs; both dislocation movements and phase transformation are suppressed compared to nanocrystals with coarser grains.

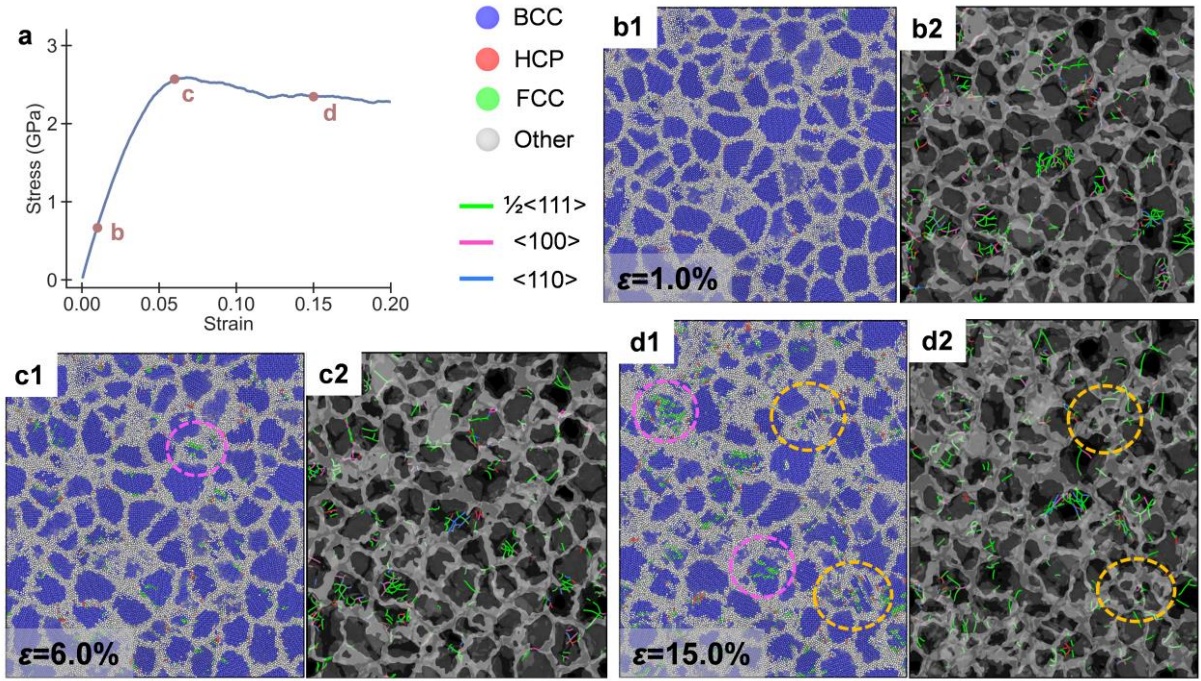

**Fig. 10** (a) The stress-strain curve of an RSS RHEA nanocrystal with $D = 4$ nm. (b1)~(d1) Atomic structures at different strains marked in (a). (b2)~(d2) The 3D-distribution of GBs, TBs and dislocations lines. Pink and yellow dashed circles mark phase transformation and deformation twins, respectively.

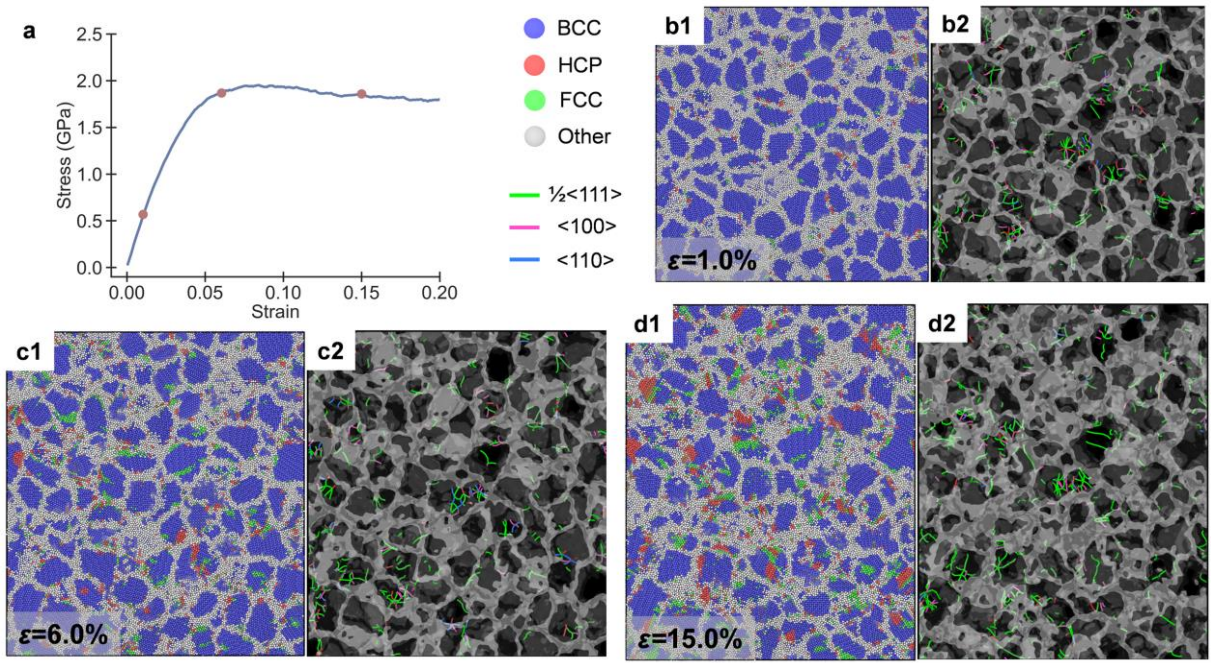

**Fig. 11** (a) The stress-strain curve of an MC RHEA nanocrystal with $D = 4$ nm. (b1)~(d1) Atomic structures at different strains marked in (a). (b2)~(d2) The 3D-distribution of GBs, TBs and dislocations lines.

The observations in Figs. 6~11 clarify the mechanisms behind the transition from HP strengthening to IHP softening in nanocrystalline HfNbTaTiZr. When $D > D_c$, networks of long, entangled dislocations (Figs. 6~8) stabilize GBs and enhance their ability to store strain energy, thereby raising the stress level under loading. In this regime, the nanocrystalline RHEA can be considered as a composite material containing a soft crystalline phase and a strong GB phase [11, 50]. As grain size decrease within the HP regime, despite the slightly decreased length of dislocation lines, the volume fraction of the stronger GB phase increases, contributing to strengthening. When $D < D_c$, dislocation starvation occurs within GBs (Figs. 9~11); the dislocation line segments are too short to entangle and form a network. As a result, the nanocrystal is regarded as a composite of a stronger crystalline phase and a soft GB phase with liquid-like viscous behavior. This explains the transition to IHP softening.

## 4. Discussions

This section presents an in-depth discussion on the size-dependent tensile response of nanocrystalline RHEA. Section 4.1 analyzes the RSS and CSRO effects on various plastic deformation mechanisms from an energetic perspective. Section 4.2 proposes a theoretical model to predict the critical grain size $D_c$ and describes the size-dependent yield strength. Finally, section 4.3 develops an analytical model to quantitatively separate the contributions of different deformation mechanisms.

### 4.1 RSS and CSRO effects on plastic deformation mechanisms

The dominant plastic mechanisms of MA, RSS and MC nanocrystals are summarized in Table 5. In MA nanocrystals, deformation twinning and intragrain dislocation slip dominate in the HP regime, while GB migration governs plastic flow in the IHP regime. In RSS nanocrystals with $D > D_c$, solid-solution effects considerably increase strength and flow stress level, while the dominant plastic mechanisms remain similar. However, in the IHP regime, the RSS effect induces a transition of GB motion from migration to sliding and amorphization. In MC nanocrystals, CSRO reduces the stress level and activates phase transformation as a new dominant mechanism in the HP regime. These findings suggest that both RSS and CSRO affect transition in the dominant plastic mechanisms. This transition is closely related to some of the core effects inherent to RHEAs: the high-entropy effect [3], lattice distortion effect [64] and cocktail effect [4-7]. Specifically, the high mixing entropy results in low Gibbs free energy and enhanced thermodynamic stability. Mismatches in atomic radius of constituent metals cause severe lattice distortion, leading to high strength and hardness. Besides, interactions between dislocations and the distorted lattice promote strain hardening and ductility. Collectively, these effects yield the so-called cocktail effect, which endows RHEAs with superior performance compared to their constituents. On the other hand, CSRO reduces both structural and chemical disorder, thereby weakening the RSS strengthening effect. In extreme cases, CSRO may cause formation of brittle intermetallic compounds [3, 64] or segregation of second phases [17, 43, 61]. While the former is detrimental to mechanical properties, the latter may improve ductility as supported by results in Fig. 5.

**Table 5** Summary of dominant plastic deformation mechanisms

| Grain size effect | Internal chemical environment | | |
|---|---|---|---|
| | MA | RSS | MC |
| HP strengthening | Deformation twinning; dislocation slip | Deformation twinning; dislocation slip | Phase transformation; deformation twin; dislocation slip |
| IHP softening | GB migration | GB sliding and amorphization | GB sliding and amorphization |

Despite previous efforts, the mechanisms underlying transitions in plastic behavior remain insufficiently understood. Here, we attempt to explain these transitions from an energetic perspective. It is commonly accepted that resistance of an alloy against dislocation slip can be represented by its generalized stacking fault energy (GSFE) curve [11, 58]. Resistance against twinning can be evaluated in a similar way [56, 57] because twins nucleate via successive glide of 1/6<111> partial dislocations on three adjacent planes (schematically shown in Fig. 12a). Correspondingly, the maximal energy per unit area is regarded as the twin fault energy $\gamma_{tf}$. The planar fault (including twinning and stacking faults) energy curves for MA, RSS and MC RHEAs are displayed in Fig. 12b. Notably, due to chemical ordering (Fig. 2), a ternary HfZrTi alloy and a binary NbTa alloy are used to represent the MC RHEA consisting of regions with similar chemical contents. As revealed by Fig. 12b, the twin fault energy $\gamma_{tf}$ is comparable to the unstable stacking fault energy $\gamma_{usf}^{(100)}$ for both MA and RSS RHEAs. The corresponding plastic mechanisms are therefore dislocation slip on (110) plane and deformation twinning on (112) plane, consistent with Table 5. Besides, the RSS effect concurrently raises both fault energies, increasing the resistance to slip and twinning and thereby material strength (Fig. 5). Fig. 12b also suggests that the Hf-Zr-Ti-rich regions in MC RHEAs tend to yield earlier owing to lower fault energies. Twinning is promoted in the Hf-Zr-Ti-rich regions, while slip on the (110) plane is more energetically favorable in the Nb-Ta-rich areas.

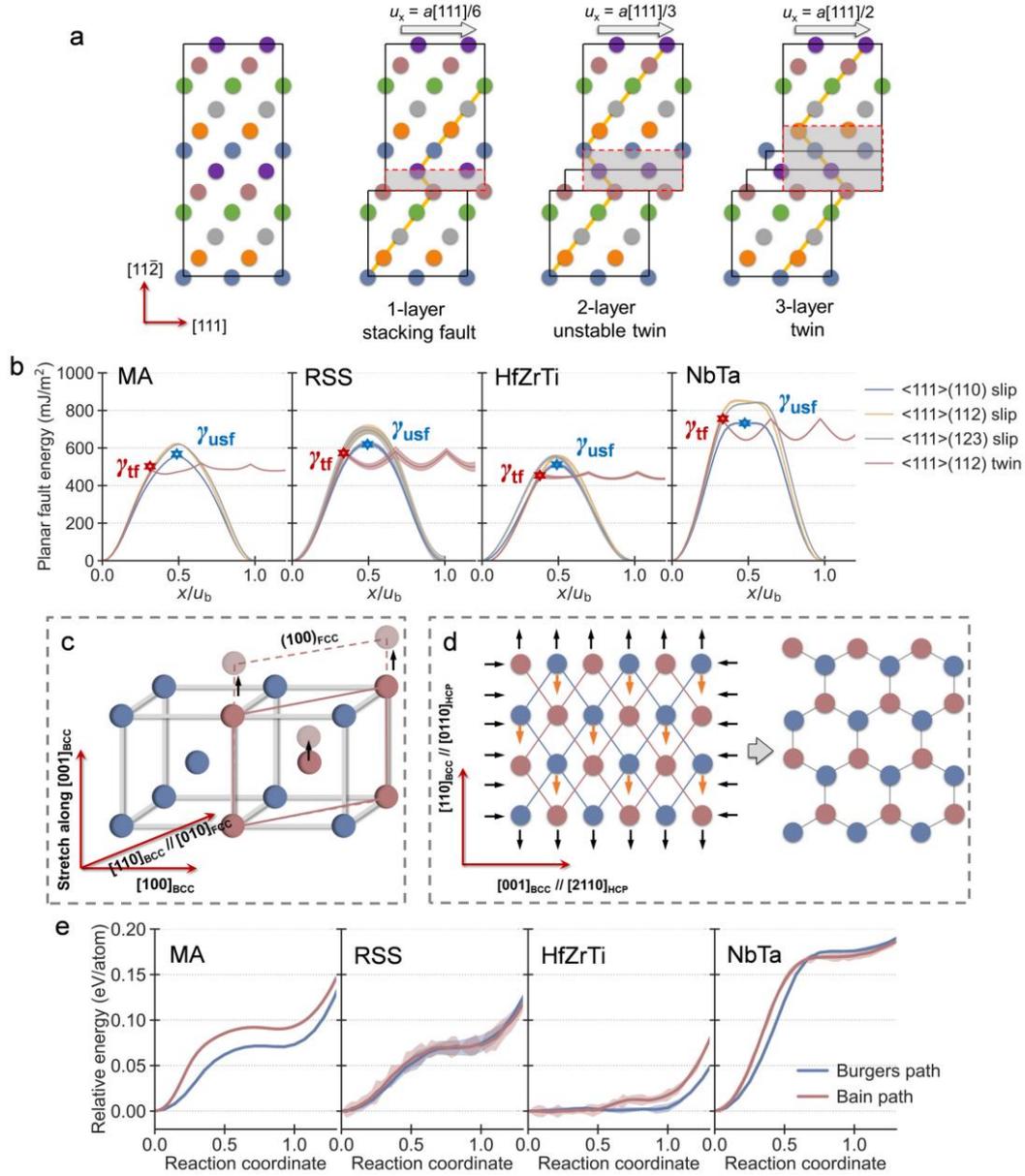

**Fig. 12** (a) The nucleation process of a deformation twin on the (112) plane, exhibiting a 3-layer growth behavior. (b) The planar fault energy curves for MA, RSS and MC RHEAs. A ternary HfZrTi alloy and a binary NbTa alloy are used to represent the MC RHEA with similar elemental distribution. (c) and (d) are the transformation paths of BCC-to-FCC (the Bain's path [36]) and BCC-to-HCP (the Burgers' path [18, 61]). (e) The energy variation during phase transformation. All energy curves are averaged over ten independent calculations.

Additionally, BCC-to-FCC and BCC-to-HCP transformation paths are depicted in Figs. 12c and 12d, respectively. The corresponding energy variations for MA, RSS and MC RHEAs are shown in Fig. 12e. The MA RHEA exhibits high transformation energy barriers. The RSS effect lowers these barriers so that FCC and HCP phases become unstable equilibrium points, which explains the reversible transformation in Fig. 7. Notably, transformation resistance is particularly low in Hf–Zr–Ti–rich regions of MC RHEAs, inducing the pronounced TRIP behavior observed in Fig. 8.

Furthermore, the transition in GB motion from migration to sliding (Figs. 9~11) can be also explained energetically. As illustrated in Fig. 13a, without the complex multi-element interactions, GBs in MA nanocrystal exhibit relatively ordered nanostructure. The kite-shaped structural units [62] induce high sliding barriers (Fig. 13b), favoring GB migration under stress. In contrast, RSS nanocrystal exhibit atomic shuffling [65, 66] at GBs (Fig. 13a), due to the multicomponent interactions. As a result, the sliding barriers decrease and GB sliding is favored. In MC nanocrystals, clustering of Hf, Zr and Ti at GBs (Fig. 8a) further lowers the sliding barrier and promotes GB sliding and amorphization.

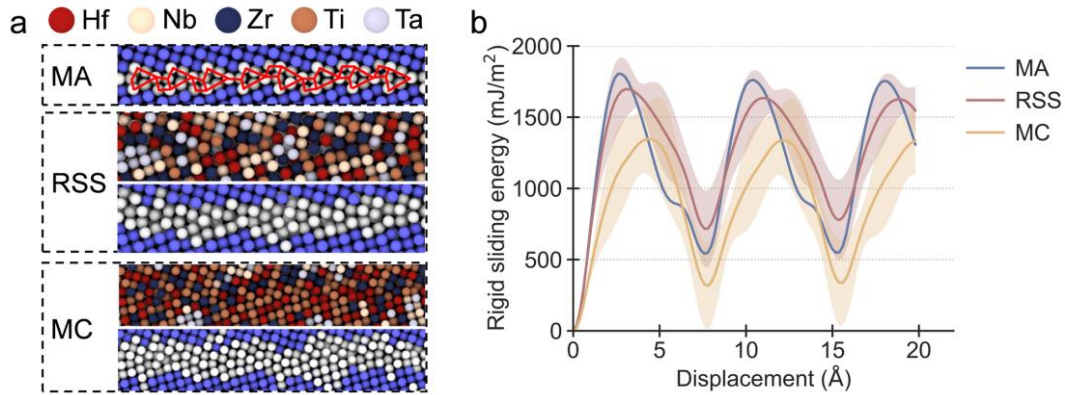

**Fig. 13** (a) The representative GB structures of the MA, RSS and MC RHEAs. (b) The averaged sliding energy curves of GBs, which is calculated via relative translation of the grains in a $\sum 5(210)$ bicrystalline model. Ten independent calculations with varying initial element distribution are performed.

### 4.2 Prediction of the critical grain size $D_c$

In this section, a theoretical model is proposed to describe the size-dependent yield strength and predict the critical grain size for MA, RSS and MC nanocrystals. According to the analysis in section 3.3, the HP–IHP transition is closely related to the relative strength of GBs. The nanocrystalline RHEA can be modeled as a two-phase composite containing amorphous GBs and crystalline grain interiors [50, 67]. When $D > D_c$, GBs contain dense, entangled dislocation networks (Figs. 6~8) and behave as the stronger phase. The plasticity is initiated by dislocation emission from GBs into grains. When $D < D_c$, dislocation starvation at GBs occurs (Figs. 9~11), GBs become the weaker phase and yielding is activated by the liquid-like viscous flow via GB motion. Therefore, HP–IHP transition can be attributed to the competition between emission of dislocations from GBs and viscous flow of the amorphous GBs. A theoretical model can be used to describe this competition as

$$\sigma_{\text{GB}}^{\text{Y}} = \min\{\sigma_{\text{H}},\ \sigma_{\text{S}}\}. \tag{4}$$

Here, $\sigma_{\text{GB}}^{\text{Y}}$ is the effective yield strength of GBs, $\sigma_{\text{H}}$ represents the hardening effect caused by intragrain dislocation emission from GBs and $\sigma_{\text{S}}$ denotes the softening effect results from GB motion. In Eq. 4, the hardening stress $\sigma_{\text{H}}$ can be expressed using the wide accepted Taylor equation as [68, 69]

$$\sigma_{\text{H}} = M\tau_{\text{H}} = M\tau_0 + M\alpha Gb\sqrt{\rho_{\text{GB}}}. \tag{5}$$

In Eq. 5, $\tau_0$ is the Peierls-like lattice friction stress[70]; $M$ and $\alpha$ are Taylor factors; $G$ is the shear modulus and $b$ is the magnitude of Burgers vector. According to previous studies[69], the GB dislocation density inversely scales with grain size: $\rho_{\text{GB}} = \frac{8m}{\pi D}$ with $m$ the ledge density. $\tau_0$ can be predicted using the following expression [71, 72]

$$\tau_0 = \frac{Gb}{a'(1-\nu)}\exp\left(-\frac{Gb}{2a'\tau_{\max}(1-\nu)}\right) \tag{6}$$

where $\nu$ is the Poisson's ratio. $a'$ is the atomic spacing along slip direction and $\tau_{\max}$ corresponds to the maximum slope of the GSFE curve along <111>(110) in Fig. 12b (i.e., the ideal shear strength). Both $a'$ and $\tau_{\max}$ are directly obtainable from MD results. Combining Eqs. 5 and 6, the hardening stress $\sigma_{\text{H}}$ as a function of $D$ can be predicted.

On the other hand, the softening stress $\sigma_S$ can be calculated by modeling activities of the amorphous GBs as viscous flow, which leads to the following expression [73, 74]

$$\sigma_S = M\tau_S = M\left(L\frac{\rho_L}{m_a}\right)\left(1 - \frac{T}{T_m}\right)f_g, \tag{7}$$

where $L$ is the heat of fusion, $\rho_L$ is the density of melted RHEA, $m_a$ is the average atomic mass, $T$ is temperature, $T_m$ is the melting point, and $f_g = [(D - \delta)/D]^3$ is the volume fraction of atoms in grain interior. Here, $\delta = 0.8$ nm is the estimated GB thickness in MD results. After calculating $\sigma_H$ and $\sigma_S$, the yield strength can be estimated using Eq. 4. Fig. 14 shows that predictions from this model align well with simulation data. The predicted $D_c$ for MA, RSS and MC nanocrystals are 14.89 nm, 13.54 nm and 10.99 nm, closely matching values in Table 4. In addition, Eqs. 5~7 also explain the variation in $D_c$ for MA, RSS and MC models. Since most material parameters are similar for three models, the key factor to the change in $D_c$ is the ideal shear strength $\tau_{max}$ (namely, slope of the GSFE curve). This implies that the stacking fault energy is decisive in the HP-IHP transition. It is also noteworthy that this model requires no fitting parameters, unlike the phenomenological HP-IHP relationship. All the related material parameters are obtained from either literature (Table S4 of the supplementary material) or simulation results. As a result, the proposed model is not only useful to analyze results of atomistic simulations, but may also be applicable in experimental and engineering fields.

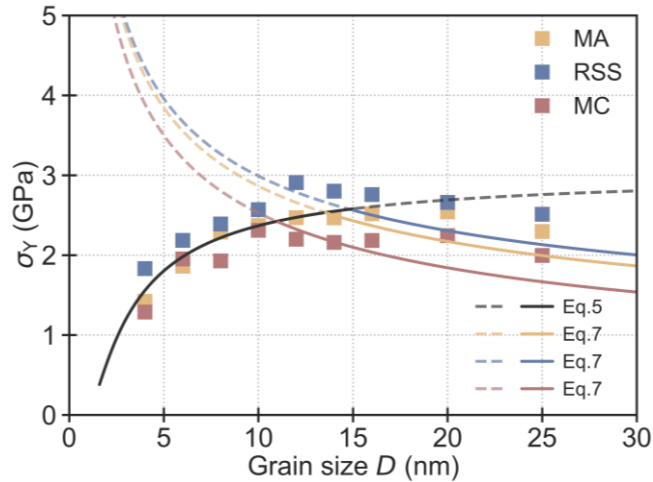

**Fig 14** The yield strengths of MA, RSS and MC nanocrystals predicted using the theoretic model.

## 4.3 Quantifying the contributions of different deformation mechanisms

As revealed in previous sections, the plastic behavior of the nanocrystalline HfNbTaTiZr is complicated due to the multiple deformation mechanisms that are activated simultaneously. For deeper insights into mechanical responses of the MA, RSS and MC nanocrystalline RHEAs, a nanostructure-based model is proposed here. Considering a HfNbTaTiZr nanocrystal subjected to uniaxial stretch, the average stress as a function of tensile strain can be decomposed according to the deformation mechanisms observed in the simulations as

$$\sigma(\varepsilon) = f_v^{BCC} [\sigma_{GB}(\varepsilon) + \sigma_{Dis}(\varepsilon) + \sigma_{Twin}(\varepsilon)] + \sigma_{Phase}(\varepsilon), \tag{8}$$

where $\sigma_{GB}$ accounts for GB responses (i.e., dislocation networks or viscous flow), $\sigma_{Dis}$ relates to intragrain dislocation movements, $\sigma_{Twin}$ results from deformation twinning and $\sigma_{Phase}$ is caused by phase transformation. $f_v^{BCC}$ is the volume fraction of BCC atoms. In Eq. 8, $\sigma_{GB}$ can be expressed as

$$\sigma_{GB}(\varepsilon) = \begin{cases} \dfrac{\sigma_{GB}^Y}{\varepsilon^Y}\varepsilon, & \varepsilon \leqslant \varepsilon^Y \\ \sigma_{GB}^Y, & \varepsilon > \varepsilon^Y \end{cases}, \tag{9}$$

$\sigma_{GB}^Y$ represents the yield strength predicted using Eq. 4 and $\varepsilon^Y$ denotes yield strain (determined using the 0.2% offset method). Eq. 9 assumes that GBs in RHEA nanocrystals can be treated as an ideal elasto-plastic phase. This assumption is reasonable due to two facts. First, the plastic flow of RHEA nanocrystals with $D > D_c$ is achieved mainly though intragranular behaviors (slip, twinning and transformation as displayed in Figs. 6~8) after the GB dislocation network reaching its maximal load capacity. Second, although the migration and sliding of GBs may also be prominent in nanocrystalline RHEAs with $D < D_c$ (Figs. 9~11), the resistance against these GB activities is relatively weak and can be approximately regarded as liquid-like viscous flow [74] as described by Eq. 7.

In addition, the stress related to intragranular dislocation slip can be described using a modified Taylor equation

$$\sigma_{\text{Dis}}(\varepsilon) = M\alpha Gb\sqrt{\rho(\varepsilon) - \rho_{\text{GB}}}, \tag{10}$$

where $\rho$ is the intragrain dislocation density during the flow stage and $\rho_{\text{GB}}$ is the density of GB dislocation network. It is noteworthy that different from most continuum crystal plasticity models that predict the evolution of dislocation density $\rho$ via analytic relations with plastic strain $\varepsilon_p$ (e.g, the Kocks–Mecking model [24, 75]), in this study $\rho$ can be simply obtained from the MD results.

Moreover, we assume that $\sigma_{\text{Twin}}$ can be depicted by a typical power-law relation [76, 77] as

$$\sigma_{\text{Twin}}(\varepsilon) = A\gamma_{\text{TB}}^n, \tag{11}$$

where $A$ and $n$ are adjustable parameters dependent on material; $\gamma_{\text{TB}}$ is the average shear strain of atoms belonging to TBs. Finally, $\sigma_{\text{Phase}}$ resulted from phase transformation is given as

$$\sigma_{\text{Phase}}(\varepsilon) = f_v^{\text{FCC}} E^{\text{FCC}}(\varepsilon - \varepsilon^Y) + f_v^{\text{HCP}} E^{\text{HCP}}(\varepsilon - \varepsilon^Y). \tag{12}$$

In Eq. 12, $f$ corresponds to the volume fractions of atoms in FCC and HCP phases; $E$ denotes the modulus of the two phases. This equation assumes elastic deformation in the newly formed phases because no apparent plastic flow is observed in them during simulations.

Substituting Eqs. 9~12 into Eq. 8, the stress-strain response of nanocrystalline RHEAs can be predicted. In this derived model, all the nanostructural variables (e.g., dislocation density $\rho$, shear strain at TBs $\gamma_{\text{TB}}$, volume fraction of phases $f$, etc.) can be obtained from MD data. Other material dependent parameters are gathered from previous literature as listed in Table S4 of the supplementary material. Fig. 15 shows that our model not only correctly predicts the mechanical response, but also separates and quantifies the contributions of various deformation mechanisms. For an MA nanocrystal with $D = 20$ nm (Fig. 15a), the internal stress is mainly carried by GB dislocation network during elastic stage. The subsequent flow stage is dominated by intragrain dislocation movements and deformation twinning. For RSS nanocrystal with an identical grain size (Fig. 15b), the contributions of intragrain dislocations and deformation twinning significantly rises, leading to an elevated stress level. For the MC nanocrystal (Fig. 15c), contributions from newly formed FCC and HCP phases become prominent at late flow stage. These phenomena are consistent with MD results in Figs. 6~8. On the other hand, when $D = 4$ nm, the total stress is mainly contributed by GB activities for MA and MC nanocrystals (Figs. 15d and 15f), while dislocation slip and deformation twinning are more active in RSS nanocrystal so that the stress level is higher. These results also agree well with Figs. 9~11. Moreover, it is noteworthy that our model is more accurate in predicting stress at elastic and early plastic stages. In fact, the reduced accuracy at later flow stage is attributed to the mathematic expressions of $\sigma_{GB}$, $\sigma_{Dis}$, $\sigma_{Twin}$ and $\sigma_{Phase}$ in Eqs. 9~12, which always give non-negative values. Therefore, this model is more accurate in describing strain hardening (i.e., $d\sigma/d\varepsilon \geq 0$) compared with stress relaxation. This issue can be addressed by introducing the effect of damage accumulation in our future study. Overall, this model is indeed effective in research on the mechanical response of RHEAs exhibiting complex plastic mechanisms.

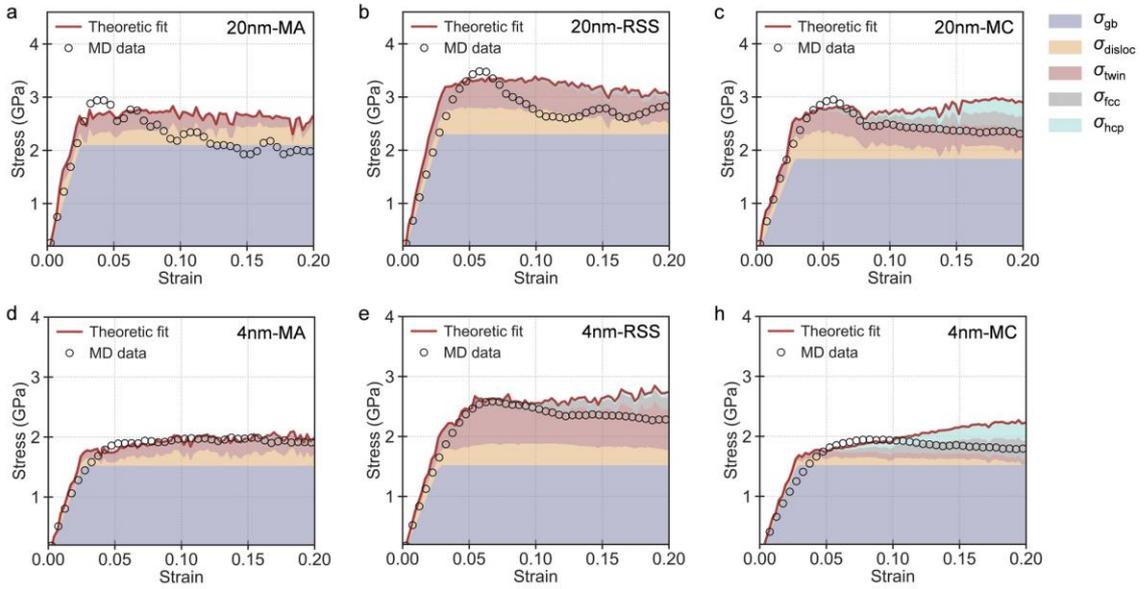

**Fig. 15** Predicted stress-strain response of MA, RSS and MC nanocrystals with (a)~(c) $D = 20$ nm and (d)~(f) $D = 4$ nm using Eq. 8. The contributions of various plastic mechanisms are separately quantified.

## 5. Conclusions

In this study, we investigate the effects of random solid-solution (RSS) and chemical short-range order (CSRO) on the size-dependent mechanical response of the HfNbTaTiZr refractory high-entropy alloy (RHEA) under uniaxial tensile loading. A machine learning (ML) framework is developed to accelerate the parameterization of interatomic force fields (FFs) for the RHEA system. The mechanical properties and nanostructural evolution of meta-atom (MA), RSS and Monte-Carlo (MC) nanocrystalline configurations are systematically analyzed. Based on the simulation data, theoretical models are proposed to describe the size-dependent yield strength and predict the critical grain size. In addition, the contributions of various plastic mechanisms to the stress-strain response are quantitatively separated. The main conclusions are summarized as follows:

1. The ML workflow considerably accelerates the development of an alloy FF (with all five metallic elements) and an MA FF (containing a single hypothetical element with averaged RHEA properties). Both FFs accurately reproduce basic RHEA properties (e.g., lattice constant, cohesive energy and elastic constants) consistent with first-principles data. Additionally, the properties of HfNbTaTiZr can be described using the rule of mixtures based on properties of its constituent metals.

2. The RSS effect markedly enhances the elastic modulus, yield strength, ultimate strength and flow stress of the nanocrystalline RHEA. The CSRO effect causes a reduction in these properties, but enhances strain-hardening ability and failure resistance. A transition from Hall-Petch (HP) strengthening to inverse Hall-Petch (IHP) softening is observed; CSRO can suppress this transition, thereby reducing the critical grain size.

3. The HP-IHP transition is ascribed to changes in nanoscale plastic mechanisms. In MA nanocrystals, deformation twinning and intragrain dislocation slip dominate in the HP regime, whereas grain boundary (GB) migration governs plastic flow in the IHP regime. The RSS effect not only improves strength by promoting dislocation slip and deformation twinning in the HP regime, but also facilitates GB sliding instead of migration in the IHP regime. CSRO favors phase transformation as a new dominant mechanism, reducing flow stress but enhancing ductility. These transitions in plastic behaviors can be explained from an energetic perspective via analyzing planar fault energies, phase transformation barriers, and GB sliding energies.

4. The size-dependent yield strengths and the critical grain sizes of MA, RSS and MC nanocrystals can be predicted using a theoretical model, which captures the competition between strengthening due to GB dislocation networks and softening caused by GB viscous flow. The RSS and CSRO effects on the critical grain size are primarily manifested by variations in the ideal shear strengths (i.e., the maximum slope of the stacking energy curves). Furthermore, by separately formulating the stress contributions from GBs, intragrain dislocations, twins and secondary phases, the individual roles of different plastic mechanisms in overall tensile response can be quantified.

**Supplementary material**

The supplementary material is available on the xxxx website.

**Acknowledgements**

The authors acknowledge the support of NSFC (Grant No: and), and the Computing Center in Xi'an.**References**